\documentclass[authoryear,preprint,review,12pt]{elsarticle}

\usepackage{graphicx}

\usepackage{amssymb}
\usepackage{amsmath}
\usepackage{lineno}
\usepackage{numcompress}
\usepackage{pdflscape}

\journal{J. Atm. Sol-Ter. Phys.,}

\begin{document}

\graphicspath{{FIG_eps/}}

\begin{frontmatter}



\title{Study of local regularities in the solar wind data and ground magnetograms.}


\author{Virginia Klausner\corref{cor}\fnref{footnote2}}
\address{DGE/CEA/National Institute for Space Research - phone: +55 12 32086819/fax: +55 12 32086810 - INPE 12227-010 S\~ao Jos\'e dos Campos, SP, Brazil.}
\cortext[cor]{Corresponding author}
\fntext[footnote2]{National Observatory - ON 20921-400, RJ, Brazil.}
\ead{virginia@dge.inpe.br}

\author{Arian Ojeda Gonz\'alez}
\address{DGE/CEA/National Institute for Space Research - phone: +55 12 32087854/fax: +55 12 32086810 - INPE 12227-010 S\~ao Jos\'e dos Campos, SP, Brazil}
\ead{arian@dge.inpe.br}

\author{Margarete Oliveira Domingues}
\address{LAC/CTE/National Institute for Space Research - phone: +55 12 32086542/fax: +55 12 32086375 - INPE 12227-010 S\~ao Jos\'e dos Campos, SP, Brazil}
\ead{mo.domingues@lac.inpe.br}

\author{Odim Mendes}
\address{DGE/CEA/National Institute for Space Research - phone: +55 12 32087854/fax: +55 12 32086810 - INPE 12227-010 S\~ao Jos\'e dos Campos, SP, Brazil}
\ead{odim@dge.inpe.br}

\author{Andres Reinaldo Rodriguez Papa\fnref{footnote5}}
\address{National Observatory - phone: +55 21 35049142/fax: +55 21 25807081 - ON 20921-400, RJ, Brazil}
\fntext[footnote5]{State University of Rio de Janeiro - UERJ 20550-900, RJ, Brazil}
\ead{papa@on.br}

\begin{abstract}

Interplanetary coronal mass ejections (ICMEs) can reach the Earth's magnetosphere causing magnetic disturbances.
It can be measured by satellite and ground-based magnetometers.
Data from the ACE satellite and from the geomagnetic field was explored here via discrete wavelet transform (DWT).
The increase of wavelet coefficient amplitudes of the solar wind parameters and geomagnetic field data analysis are well-correlated with the arrival of the shock and sheath region.
As an auxiliary tool to verify the disturbed magnetic fields identified by the DWT, we developed a new approach called effectiveness wavelet coefficient (EWC) methodology.
The first interpretation of the results suggests that DWT and EWC can be effectively used to characterize the fluctuations on the solar wind parameters and its contributions to the geomagnetic field.
Further, these techniques could be implemented to real-time analysis for forecast space weather scenarios.

\end{abstract}


\end{frontmatter}
\linenumbers


\section{Introduction}
\label{Introduction}
Several years ago, in 1957, Hannes Alfv\'en postulated that the solar wind was magnetized and that the solar-wind flow draped the magnetic field over comets, forming a long magnetic tail downstream in the antisolar direction \citep{KiveltonRussel1995}.
So, this solar plasma expands out from the Sun driven by thermo-electrodynamical processes.
The solar magnetic field propagates ``frozen'' in the solar wind in a spiral-like configuration due to the Sun's rotation.
The geomagnetic field can be considered as a dipole magnetic field, limited by the changes in the solar wind density and velocity, and by the variation in the strength and orientation of the interplanetary magnetic field (IMF) \citep{Tascione1988}.
The electrodynamical interaction between that solar plasma and the Earth's magnetized atmosphere generates a complicated interrelated current system such as magnetosphere current, tail current, ring current, field-aligned current and ionospheric current, adding to a lower electric circuit with atmospheric discharges \citep{Mendes2005}.
In other words, the interaction between the solar wind and IMF with the magnetosphere-ionosphere-ground creates a great variety of complex processes, which generate geomagnetic activity \citep{Mendes2006}.

During solar events, when the solar plasma in expansion incides upon the Earth's intrinsic magnetic field, a substantial transfer of energy into the terrestrial magnetosphere may take place.
Then, the normally existing magnetospheric and ionospheric quiet currents are widened and intensified.

Among other phenomena, one of the characteristic signature is the geomagnetic storm, a depression in the horizontal component of the Earth's magnetic field (H) at middle to low latitude.
The key parameters that control the solar wind magnetospheric coupling are the direction, the strength and the duration of the interplanetary magnetic field (IMF).
For example, intense magnetic storms ($Dst\leq -100 \: nT$) are caused by IMF southward component stronger than 10 nT at least for 3 hours \citep[e.g.,][]{Gonzalez1987}.
Solar wind speed and density also play an important role in the formation of the ring current, though their exact role is still controversial \citep[e.g.,][]{Huttunen2005,Wang2003,Fenrich1998,Gonzalez1987}.

An important solar event is the coronal mass ejection (CME) because it can cause geomagnetic storms.
They are observed near 1 AU and are called interplanetary coronal mass ejections (ICMEs).
The term magnetic cloud (MC) is used to characterize an ICME having a specific configuration of IMF and plasma density \citep[e.g.,][]{Gosling1990,KleinBurlaga1982,Burlaga1981}.
The MCs have values of plasma beta significantly lower than 1 which is the ratio of plasma pressure to magnetic pressure.
Near 1 AU, MCs have enormous radial sizes (0.28 AU), with an average duration of 27 h, an average peak magnetic field strength of 18 nT and the average solar wind speed 420 km/s \citep[e.g.,][]{Goldstein1983,Lepping2000,KleinBurlaga1982}.
In \citet{Goldstein1983}, it was suggested for the first time that MCs are force-free magnetic field configurations (that is, when $\nabla\times \vec B=\alpha (r)\vec B $).
The constant $\alpha$ solution for a cylindrical symmetric force-free equation was given by \citet{Lundquist1950}. 

In this paper, we work with time series variations of solar wind parameters due to solar events to verify its behaviors before reaching the Earth, the ``cause'',
and the ``effect'' of the solar wind-magnetosphere interaction, the geomagnetic storm, in the ground magnetograms.
Several magnetic storms that occurred on  April, 2001 have been analyzed.
These magnetic storms were associated with magnetic clouds occurrence although shocks have also been observed.
According to \citet{Farrugiaetal:1993}, magnetic clouds can be used for the study of the solar wind energy input to the magnetosphere since the IMF components vary smoothly with time and retain the polarity for relatively long intervals.
The discrete wavelet transform (DWT) have been selected and used with three levels of decomposition in order to detect the small regularities in the solar parameters and the geomagnetic data.
The rest of the paper is organized as follows: 
Section \ref{Geomagnetics Storm} is devoted to present a brief introduction of geomagnetic storms, showing an example of solar-interplanetary-magnetosphere coupling.
Section \ref{Dataset and Methodology}, the data and the analyzed period is presented. 
Section \ref{Results and Discussion}, gives a discussion about the results. Finally, section \ref{Final Remarks} presents the conclusion of this work.

\section{Geomagnetics Storm}
\label{Geomagnetics Storm}

The primary causes of geomagnetic storms are supposed to be strong dawn-to-dusk electric fields associated with the passage of southward directed interplanetary magnetic fields,
Bs, passing the Earth for sufficiently long intervals of time.
The solar wind energy transfer mechanism is magnetic reconnection between the IMF and the Earth's magnetic field \citep{Gonzalez1994}.

As consequence, the level of magnetosphere activity varies widely.
Geomagnetic activity is classified by intensity and usually described by the variation of indices to distinguish between a quiet and an active day (occurrence of storm or substorm).
The index most used in order to quantify the effects on low latitudes is the Dst index, and recently, Sym-H.
It represents the variations of the H component due to changes of the ring current \citep{Tascione1988}.
The Sym-H is essentially the same as the traditional hourly Dst index.
The main characteristic of the $1$ minute time resolution Sym-H index is that the solar wind dynamic pressure variation are more clearly seen than indices with lower time resolution.
Its calculation is based on magnetic data provided by eleven stations of low and medium latitude.
Only 6 of the stations are used for its calculation of each month, some stations can be replaced by others depending on the data conditions.

The principal defining property of a magnetic storm is the creation of an enhanced ring current due to the increase of the trapped magnetospheric particle population.
These particles present a drift due to magnetic field curvature and gradient, which leads the ions to move from midnight to dusk and electrons from midnight toward dawn surrounding the Earth close to the dip equator \citep{Gonzalez1994}.

By the time of the late recovery phase, when the ring current is symmetric and trapped on closed trajectories most of its energy has already been dissipated.
The ring current becomes symmetric just after the minimum Dst is reached \citep{Daglis2002}.
The asymmetric structure (during the main phase) and the symmetric structure (during the recovery phase) were observed by \citet{EbiharaEjiri2000} using data
obtained by the polar orbiting satellite NOAA 12.

It must be noticed that the quasi-trapped particles in the inner radiation belt can sink to the South Atlantic Magnetic Anomaly (SAMA)
which is characterized by a global minimum in the Earth's total magnetic field intensity \citep{Pintoetall1992}.
These particles can reach ionospheric heights.
\citet{Nishinoetall2002} verified variation on the ionospheric parameters measured by ionosonde at Cachoeira Paulista (Brazil).
Also, the processes which cause energetic electrons to precipitate in the atmosphere are:
magnetospheric wave-particle interaction, lightning or artificially induced wave-particle interaction, drift-resonance interactions
and wave-particle interactions generated by plasma instabilities \citep{PintoGonzalez1989}.
In this paper, the Vassouras station was chosen to study the geomagnetic variations over the Brazilian sector because taking into account somewhat the SAMA influence.

\section{Dataset and Methodology}
\label{Dataset and Methodology}

In this section, we will present the ACE satellites data used to characterize the variations on the solar parameters due to the ICMEs propagation.
Also, we will present the data used to analyze the effects of the magnetospheric activity effects generated on the ground.
Our study concerns to magnetic events that occurred on April, 2001.
These solar events occurred during the solar maximum of the 23th solar cycle.
The method used here is based on the Discrete Wavelet Transform (DWT).
It will be used to verify the increase of wavelet coefficient amplitudes associated to the ICMEs propagation and its effectiveness in a development of geomagnetic storms.

\subsection{ACE satellite data}

In this paper, the geomagnetic activity during the month of April, 2001 was studied using ACE satellite dataset and Sym-H index.
Both of them are available at the NOAA web site \citep{Spidr2008}.

Fig.~\ref{fig:ACE}(a) shows the behavior of the mean values of solar wind parameters and the Sym-H index.
Each panel presents, from top to bottom, IMF components (Bx, By, Bz in GSE coordinate), plasma density, velocity (Vx component) and Sym-H index.
Our interest is to characterize the variations on the solar parameters due to the ICMEs propagation.
Also, we used the Sym-H index to verify the global magnetic field reductions during storms due to ICMEs propagation on April 2001.

All time series represent ACE and geophysical parameters with $1$ minute resolution.
On April $2001$, it happened several solar disturbances which were detected by ACE instruments.
Some ICMEs that caused these geomagnetic disturbances were studied by \citet{WuLepping2003, Huttunen2005}. 

\cite{Huttunen2005} used ACE satellites data to investigate which were the possible candidates to MCs events.
In this paper, we will verify the increase of wavelet coefficient amplitudes associated to the same ICMEs events studied by \cite{Huttunen2005}.
Table~\ref{tabela:MCList} shows three magnetic clouds (MCs) identified by \citet{Huttunen2005} in each one is presented the MC arrival and stop time, the inferred flux-rope type and Dst index (day and time (UT) of occurrence).
The minimum Dst index is important to characterize the MCs geoeffectiveness.
According to \citet{Zhang1988,BothmerSep2003} the geomagnetic response of a certain MC depends greatly on its flux-rope structure (Table~\ref{tabela:MCList}, column 4).

\begin{table}[ht]
 \caption{Magnetic Clouds identified by ACE data on April, $2001$ \citep{Huttunen2005}.
The columns from the left to the right give: Shock date (day and hour (UT)), MC start date (day and hour (UT)), MC end date (day and hour (UT)), inferred flux-rope type \citep[e.g.,][]{Huttunen2005}, the minimum value of the Dst index, it date of the previous one (day and hour (UT), if the sheath caused the storm, it is indicated by ``sh'')}.
\centering
\scriptsize{
\begin{tabular}{ c c c c c c}
 \hline
 Shock     & MC, Start  & MC, Stop  & type& $Dst_{min}$ & Day and time (UT) of $Dst_{min}$ \\ [0.5ex]
\hline
11, 15:18 & 12, 10:00 & 13, 06:00 & WNE & sh(-259)  & 11, 23:00      \\
21, 15:06 & 21, 23:00 & 22, 24:00 & WSE & -103      & 22, 15:00      \\
28, 04:31 & 29, 00:00 & 29, 13:00 & SEN & -33       & 29, 03:00      \\[1ex]
\hline
\end{tabular}
}
\label{tabela:MCList}
\end{table}

\subsection{Ground magnetics data}

The magnetic stations considered in the analysis are  Honolulu (HON), San Juan (SJG), Vassouras (VSS), Hermanus (HER) and Kakioka (KAK).
We are interested to examine how the effects within the magnetic storms on April, 2001 can affect the global space and time configuration of the ring current at low-latitudes and the additional variations of the H-component related to the effect of equatorwards penetration of electric fields from the field-aligned current \citep{WuLiou2004}.
These selected magnetic stations are used to calculated the Dst index, excluding VSS, because the Dst index reflects the magnetic variations related to changes of the ring current.
We choose the ground magnetic observatory of VSS station because we are also interested to understand better how these processes can effect the magnetic response to disturbed times in the SAMA.

Another point of interest is the correlation of the response of these chosen stations due to the development of geomagnetic storms.
The geographic and geomagnetic localization of these stations are shown in Table~\ref{table:CoordStation}.

\begin{table}[ht]
 \caption{INTERMAGNET network of geomagnetic stations used in this study.}
\centering
\scriptsize{
\begin{tabular}{l l r r r r}
\hline
Station& IAGA code & \multicolumn{2}{c}{Geographic coord.} & \multicolumn{2}{c}{Geomagnetic coord.}\\
\cline{2-6}
       &   & Lat.($^o$) & Long.($^o$) & Lat.($^o$) & Long.($^o$)  \\[0.5ex]
\hline
Honolulu (United States) &HON     &21.32      &-158.00       &21.59      &-89.70   \\
San Juan (Puerto Rico) &SJG     &18.12      &-66.15        &27.93      &6.53  \\
Vassouras (Brazil) &VSS     & -22.40    & -43.65       &-13.43     &27.06 \\
Hermanus (South Africa) &HER     &-34.41     &19.23         &-33.89     &84.68   \\
Kakioka (Japan) &KAK     & 36.23     & 140.18       & 27.46     &-150.78\\[1ex]
\hline
\end{tabular}
}
\\Source: http://wdc.kugi.kyoto-u.ac.jp/igrf/gggm/index.html (2010)
\label{table:CoordStation}
\end{table}

Fig.~\ref{fig:Ground}(a) shows the behavior of the H-component of the Earth's geomagnetic field and the Sym-H index detected on April 2001.
Each panel presents, from top to bottom, the H-component obtained at HON, SJG, VSS, HER and KAK; and the Sym-H index.
We could observe a magnetic signature of few geomagnetic storms with different magnitudes.
The geomagnetic storms can consist of four phases: sudden commencement, initial phase, main phase and recovery phase \citep{Gonzalez1994}.
The four major geomagnetic storm which present all four phases are: a very-intense magnetic storm with minimum $Dst = -271\;nT$ at 24:00 UT on April 11, 2001, a second one started on April 18, 2001 with minimum $Dst = -114\;nT$ at 08:00 UT, followed by the third one started on April 21, 2001 with minimum $Dst = -102\;nT$ and a last storm occurred on April 28, 2001 with minimum $Dst = - 47\;nT$.

\subsection{Discrete Wavelet Transform (DWT)}

The discrete wavelet transform (DWT) has the following propriety: the larger amplitudes of the wavelet coefficients are associated with locally abrupt signal changes or ``details'' of higher frequency. 
In the work of \cite{MendesMag2005} and the following work of \cite{MendesdaCostaetal:2011}, a method for the detection of the transition region and the exactly location of this discontinuities due to geomagnetic storms was implemented. 

The DWT produces the so called wavelet coefficients at different levels and it is proved that their amplitudes can be used to study the local regularity of the analyzed data \citep{mallat1997}.
As smaller is the amplitude as regular is the analyzed data. 
Therefore, where the amplitudes are large we can associate it to some disturbance on the signal \citep{MendesMag2005}.

The wavelet transform in level $j+1$ is given by
  \begin{align} 
              d_k^{j+1} &= 2 \sum\limits_m g(m-2k) \; c_m^{j},
   \end{align}
where  $g$ is a high-pass filter, $d_k^{j+1}$ is the wavelet coefficient at level $j+1$,  and $c_m^{j}$ are the scale coefficients at level $j$.
In this transform,
\begin{align} 
       c_k^{j+1} &= 2 \sum\limits_m h(m-2k) \; c_m^{j},
   \end{align}
and $h$ is a low-pass filter.

In this study, we considered $j=0$ as the most refine level of the multi-level decomposition which is associated to one minute data resolution.
In other word,  $c_k^{j=0}$ is the mean time fluctuation computed from the raw dataset.

In our case, the highest amplitudes of the wavelet coefficients indicate the singularities on the geomagnetic signal in association with the disturbed periods.
On the other hand,  when the magnetosphere is under quiet conditions for the geomagnetic signal, the wavelet coefficients show very small amplitudes.
In this work, we applied this methodology with Daubechies orthogonal wavelet function of order 2 (db2) with data sample rate of one minute time resolution.
Also, we used with three levels of decomposition which are associated to pseudo-periods of 3, 6 and 12 minutes.
On the physical point of view, these periods are related to the propagation of Alfv\'en wave and the Pci 5 pulsations \citep{Saito1969}.

In order to facilitated the visualization and the analysis of the wavelet coefficients, we developed a methodology called the effectiveness wavelet coefficients (EWC).
This new method has an advantaged over the DWT method presented above.
In the DWT method, we have to verified the wavelet decomposition on all the calculated chosen levels, in our case, three levels.
Only the local regularities detected on the three levels of decomposition could be considered due to a physical process, in this case, a ICMEs propagation.
The EWC simplifies the DWT method by reducing all the wavelet decomposition levels for only one level.
It corresponds to the weighted geometric mean of the square wavelet coefficients per hour.
It is accomplished by weighting the square wavelet coefficients means in each level of decomposition as following

\begin{equation}
 EWC=\frac{4\,\sum_{i=1}^{N}d1+2\,\sum_{i=1}^{N}d2+\sum_{i=1}^{N}d3}{7},
\end{equation}

where $N$ is equal to $60$ because our time series has one minute resolution.

The advantage of the EWC over the DWT is that it make easy the visualization and the analysis of the wavelet coefficients.
In this new approach, we only have to analyze one decomposition level.
The EWC is weighting according with the decomposition levels because the wavelet coefficients are calculated by filtering the original signal and by subsampling the resulting signal by a factor 2 \citep{mallat1997}.
In this case, the first level of the decomposition has half of the number of the points of the original data, the second level of the decomposition has half of the number of the points of the first level of the decomposition and so on and so forth.

\section{Results and Discussion}
\label{Results and Discussion}

In this section, we first analyzed the solar wind parameters due to solar events that happened at April, 2001 to verify it behaviors before reaching the Earth, the ``cause'' of the geomagnetic storms.
Second, we verified the ``effect'' of the solar wind-magnetosphere interaction, the geomagnetic storm, in the ground magnetograms.
Following, we examined the correlations between the ``cause'' and the ``effect''.

\subsection{ACE}

\begin{landscape}
\begin{figure}[hbt]
\centering
\scriptsize{
\begin{tabular}{cc}
(a) ACE data & (b) $(d^j)^2$ for ${j=1}$  \\
\includegraphics[width=10.5cm, height=6.5cm]{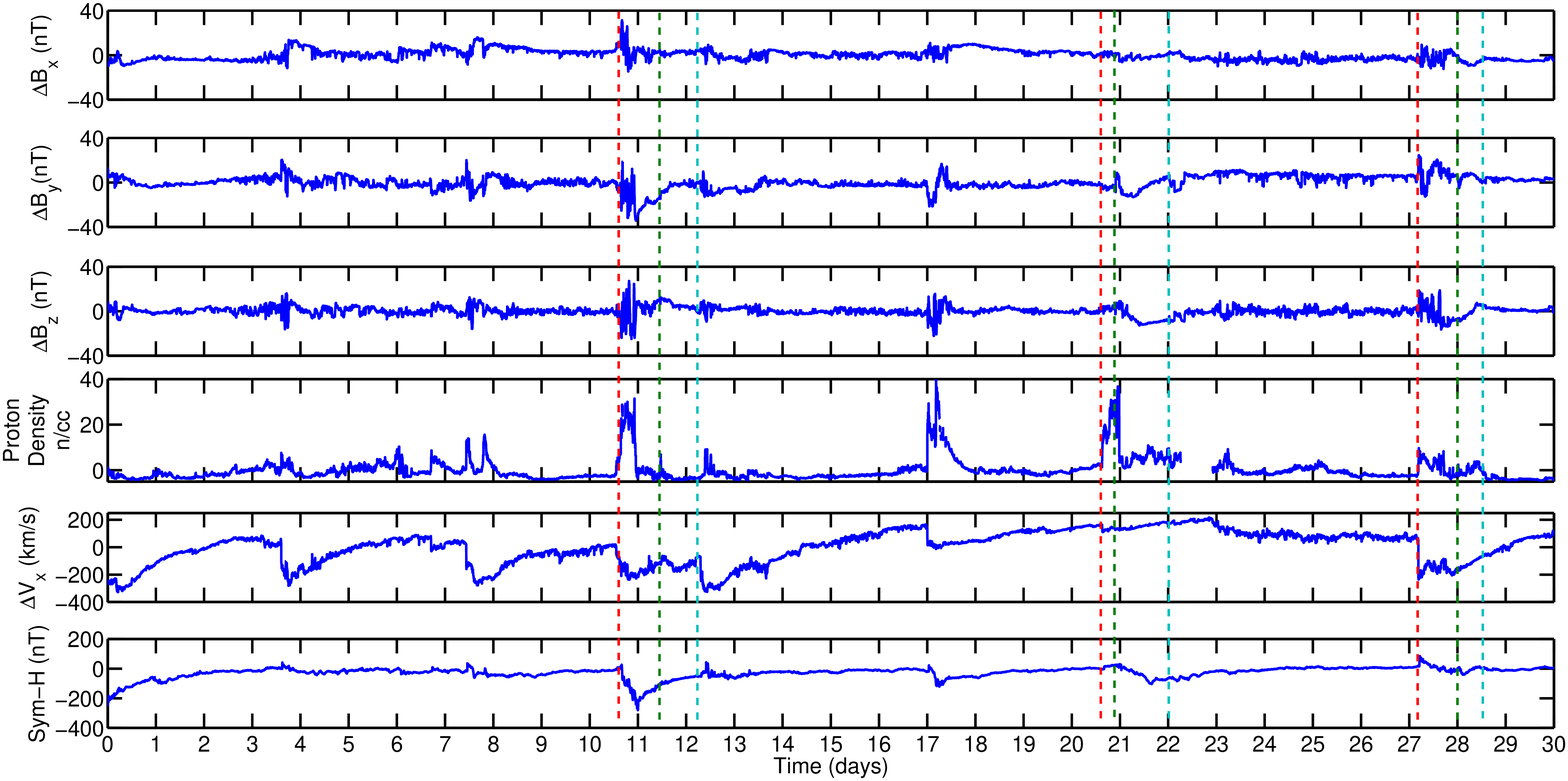} &
\includegraphics[width=10.5cm, height=6.5cm]{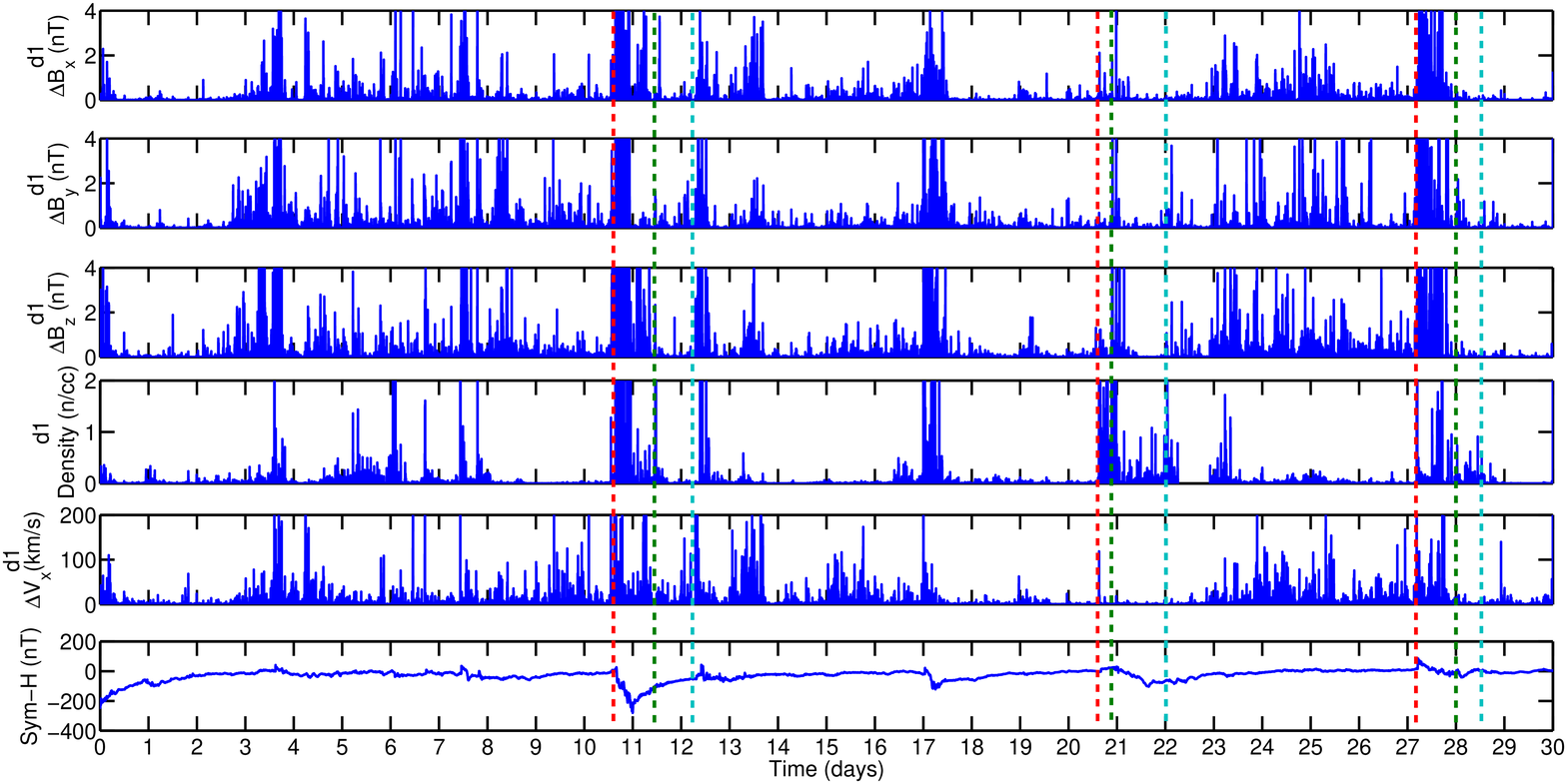}\\
(c) $(d^j)^2$ for ${j=2}$  & (d) $(d^j)^2$ for ${j=3}$ \\
\includegraphics[width=10.5cm, height=6.5cm]{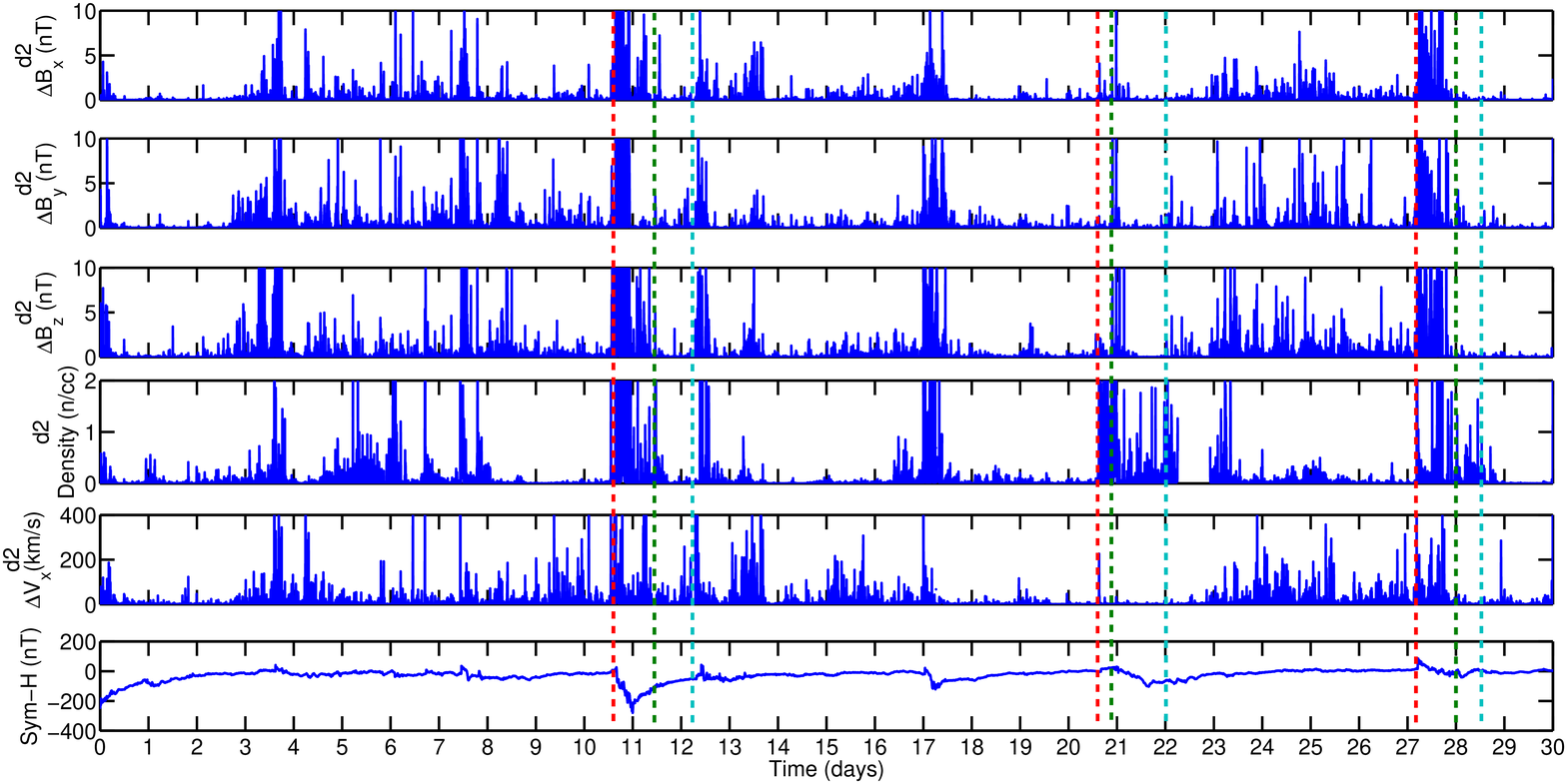} &
\includegraphics[width=10.5cm, height=6.5cm]{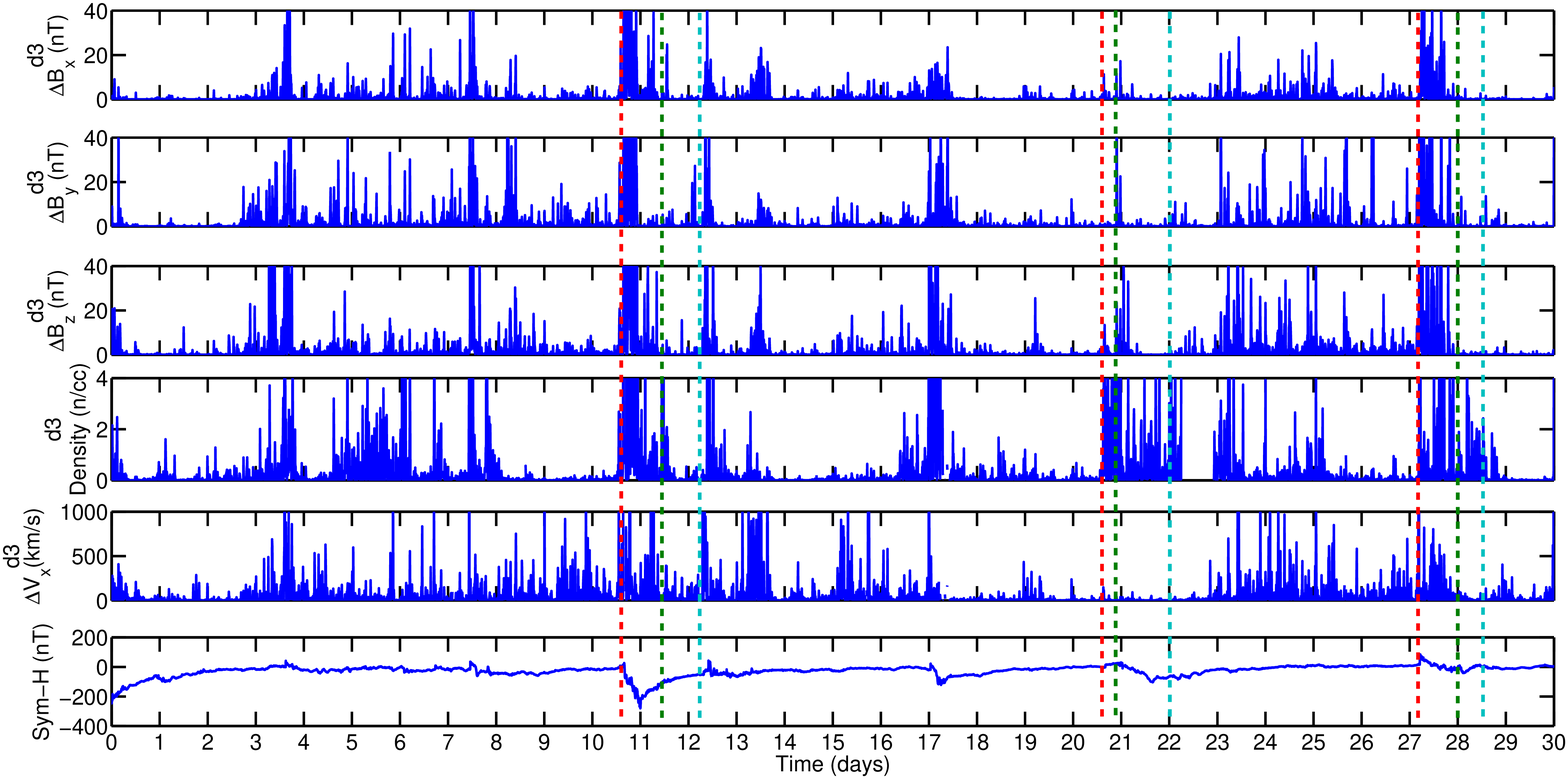}\\
\end{tabular}
}
\caption{ACE: (a) Magnetic field components (X, Y and Z), Plasma Density, Velocity (X-component) and Sym-H index; and (b,c,d) the wavelet coefficients $(d^j)^2$ for $j=1,2,3$ and Sym-H index, respectively.}
\label{fig:ACE}
\end{figure}
\end{landscape}

Fig.~\ref{fig:ACE} is divided in four panels.
At top left, it shows the behavior of the solar wind parameters (IMF components (Bx, By, Bz in GSE coordinate), plasma density and velocity (Vx)) and the Sym-H index for April, 2001.
We highlighted the three MCs events shown in Table~\ref{tabela:MCList} using the dashed lines in red for shock time, in green for the MCs arrival time and in light blue for MCs stop time.
At top right, at bottom left and at bottom right, the panels show the behavior of the square wavelet coefficients for these solar wind parameters at three first levels j = 1, 2 and 3, denoted by $d^1$, $d^2$, $d^3$, respectively.
The wavelet coefficients identified the variations of the ACE parameters due to solar wind events with different properties and degrees of disturbance.
The MCs events shown on Table~\ref{tabela:MCList} were all identified by the increase of the wavelets coefficients at the three decomposition levels.

The wavelet coefficients present an increase of amplitude during the first event shown in Table~\ref{tabela:MCList}.
During the MC of WSE-type, the magnetic field vector rotates from the west (W) at the leading edge to the east (E) at the trailing edge, being south (S).
Also, the Bz-component has the same sign during the MC propagation and it has the axis highly inclined to the ecliptic (Unipolar MCs).
The first MC event of WNE-type caused one extreme storm happened with a minimum Dst = -256 nT (shock) at 23:00 UT on 11th of April.
It was identified by wavelet coefficients at all three levels of decomposition during its shock time.
A great number of wavelet coefficients present an increase of amplitude were found between the shock and the MC starting time at all three levels of decomposition.
These wavelet structures are associated to the sheath, region of solar wind bounded by the shock front and the ICME leading edge \citep{Owensetal:2005}.

The second MC event of SNE-type also presented an increase of wavelet coefficients amplitude. 
This event caused medium degree of disturbance of Dst index according to NOAA classification (Dst = - 33 nT).
The plasma density presented wavelet structures during the shock and the sheath region at all three levels of decomposition.
This could be a consequence of the low values of plasma density associated with this particular event in the solar wind.
The third MC event with WSE flux rope type also was identified by wavelet coefficients between its shock and the MC starting time.
Its wavelet structures were very similar to the first MC.

Apart of the three MCs events identified by \cite{Huttunen2005}, the wavelet coefficients present an increase of amplitude during other periods.
This fact could be associated to the occurrence of ICMEs observed by the Solar and Heliospheric Observatory/Large Angle and Spectrometric Coronagrafic \citep{CaneRichardson:2003}.
The probable ICMEs identified by \cite{CaneRichardson:2003} are listed on Table~\ref{tabela:MCListCane}.
The columns indicate the estimated time of related disturbance in the upstream solar wind, the start, and the end times of the ICME, the quality of the estimated boundary times (1, most accurate; 3, ill-defined) and the minimum value of the geomagnetic Dst index respectively.
In the last column, ''2'' indicates whether the ICME has been reported as a MC which can be modeled by a force-free flux rope ($http://lepmfi.gsfc.nasa.gov/mfi/mag\_cloud\_pub1.html$); ''0'' indicates that the field shows little evidence of rotation, i.e., the ICME is not reported as cloud.
The other four ICMEs studied by them are not reported as MCs.
However, the three MCs reported by them are the same of the work of \cite{Huttunen2005}.

Altogether, the wavelet structures was increased during the shock and the sheath region at all three levels of decomposition for all the MC events mentioned above.

\begin{landscape}
\begin{figure}[hbt]
\includegraphics[width=21cm]{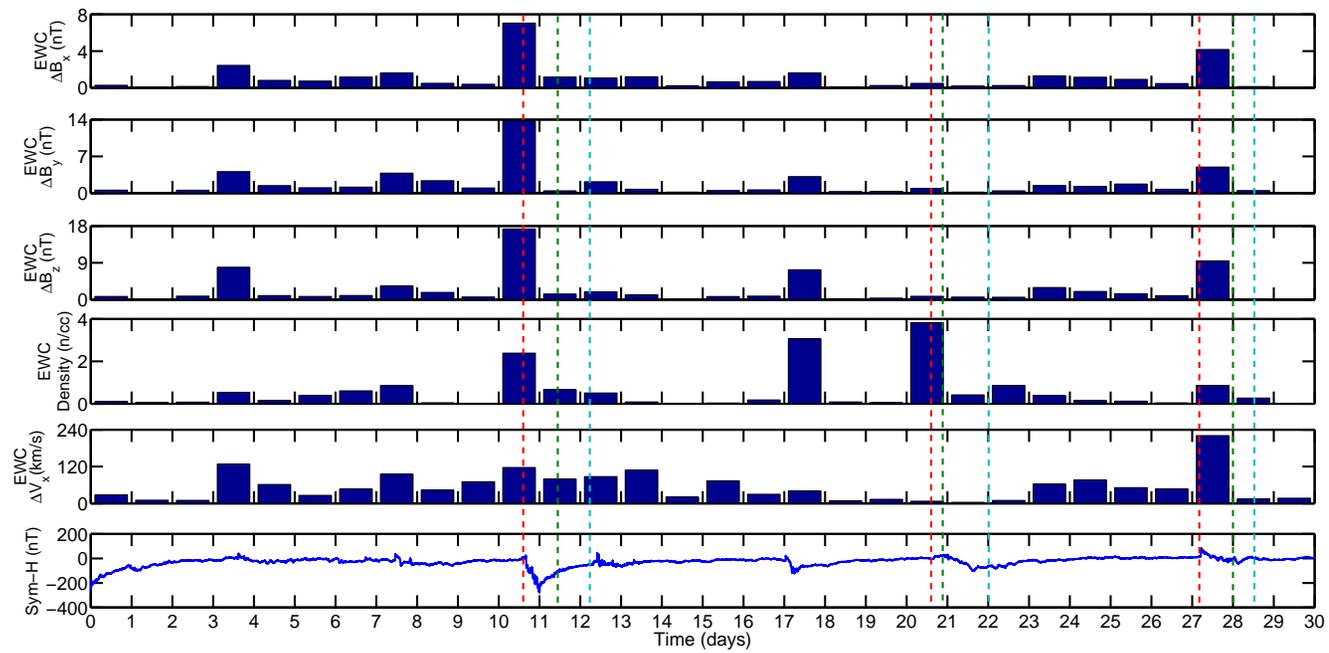}
\caption{The effectiveness wavelet coefficients on April, 2001 for the ACE dataset.}
\label{fig:EWC_ACE}
\end{figure}
\end{landscape}

Fig.~\ref{fig:EWC_ACE} shows EWCs on April, 2001 for the ACE dataset.
On the vertical axis, each panel presents the the weighted geometric mean of the square wavelet coefficients per hour, in other words, the EWCs.
And, on the horizontal axis, the time of day in Universal Time (UT).
On the analyses of the IMF components, the wavelet coefficients presented higher amplitude during the shock and the sheath region of the first and third MC.
However, the second MC was better detected by EWCs of plasma density than by the IMF components, during its shock and sheath region.
Also, there was an increase of EWCs amplitude of all solar parameter during the third shock and sheath region.
This characteristic depends of the interplanetary disturbance and the physical processes involved such as: magnetic reconnection, viscous dissipation (type interactions) or the mixture of both cases.

\begin{table}[ht]
 \caption{ICMEs identified by LASCO data on April, $2001$ \citep{CaneRichardson:2003}.
The columns from the left to the right give: Shock date (day and hour (UT)), MC start date (day and hour (UT)), MC end date (day and hour (UT)), Quality, the minimum value of the Dst index (nT) and if the the cloud was identified.}
\centering
\scriptsize{
\begin{tabular}{ c c c c c c}
 \hline
 Shock     & ICME, Start  & ICME, Stop  & Quality& $Dst_{min}$ & MC? \\ [0.5ex]
\hline
4, 14:55 & 5, 11:00 & 7, 03:00 & 3 & -38     & 0\\
8, 11:01 & 8, 19:00 & 10, 10:00 & 2 & -51    & 0\\
11, 13:43 & 11, 22:00 & 13, 07:00 & 2 & -257 & 2\\
13, 07:34 & 13, 09:00 & 14, 12:00 & 1 & -66  & 0\\
18, 00:46 & 18, 12:00 & 20, 11:00 & 2 & -100 & 0\\
21, 16:01 & 21, 23:00 & 23, 08:00 & 1 & -104 & 2\\
28, 04:31 & 28, 14:00 & 1, 02:00 & 2 & -33   & 2\\[1ex]
\hline
\end{tabular}
}
\label{tabela:MCListCane}
\end{table}


Also, it can be noticed an smaller EWCs on the IMF components that corresponded to the ICMEs propagation between the days of 4--7, 8--10, 11,13 and 18--20, see Table~\ref{tabela:MCListCane}.
Once more, the larger EWC amplitudes of these solar parameter occur during the shock and sheath region.

\subsection{Ground magnetics data analysis}

\begin{landscape}
\begin{figure}[hbt]
\centering
\scriptsize{
\begin{tabular}{cc}
(a) Ground data & (b) $(d^j)^2$ for ${j=1}$  \\
\includegraphics[width=10.5cm, height=6.5cm]{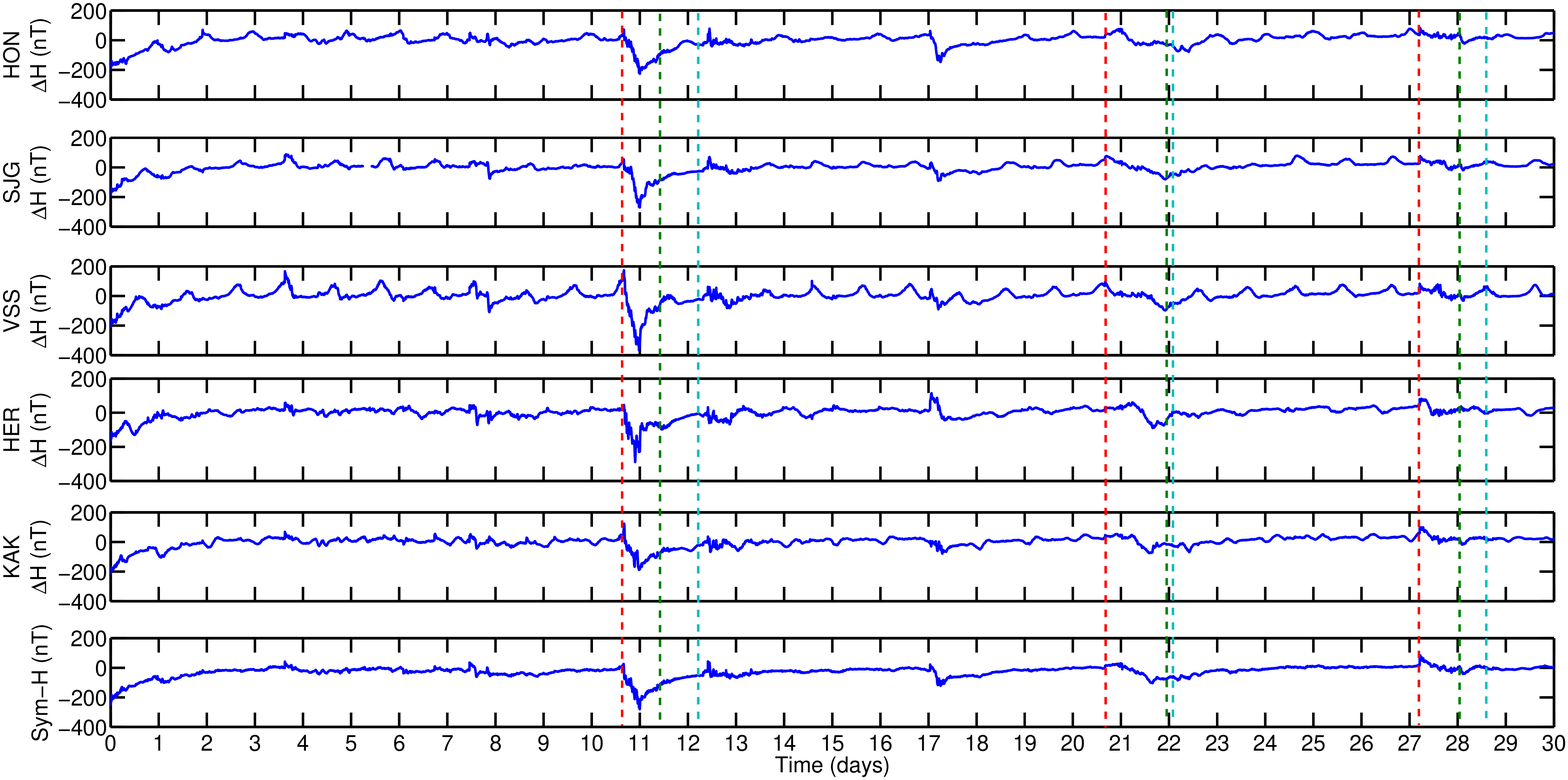} &
\includegraphics[width=10.5cm, height=6.5cm]{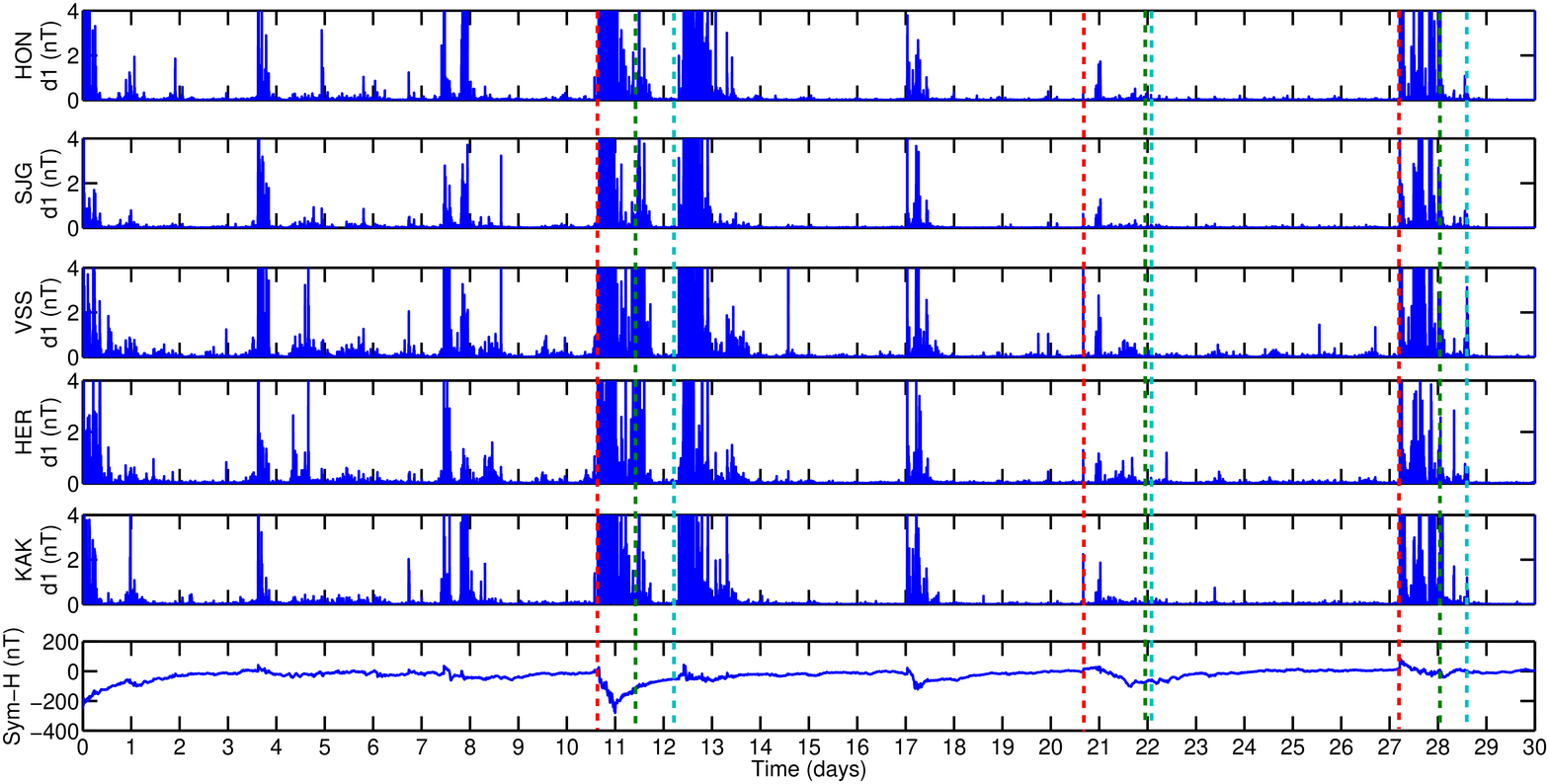}\\
(c) $(d^j)^2$ for ${j=2}$  & (d) $(d^j)^2$ for ${j=3}$  \\
\includegraphics[width=10.5cm, height=6.5cm]{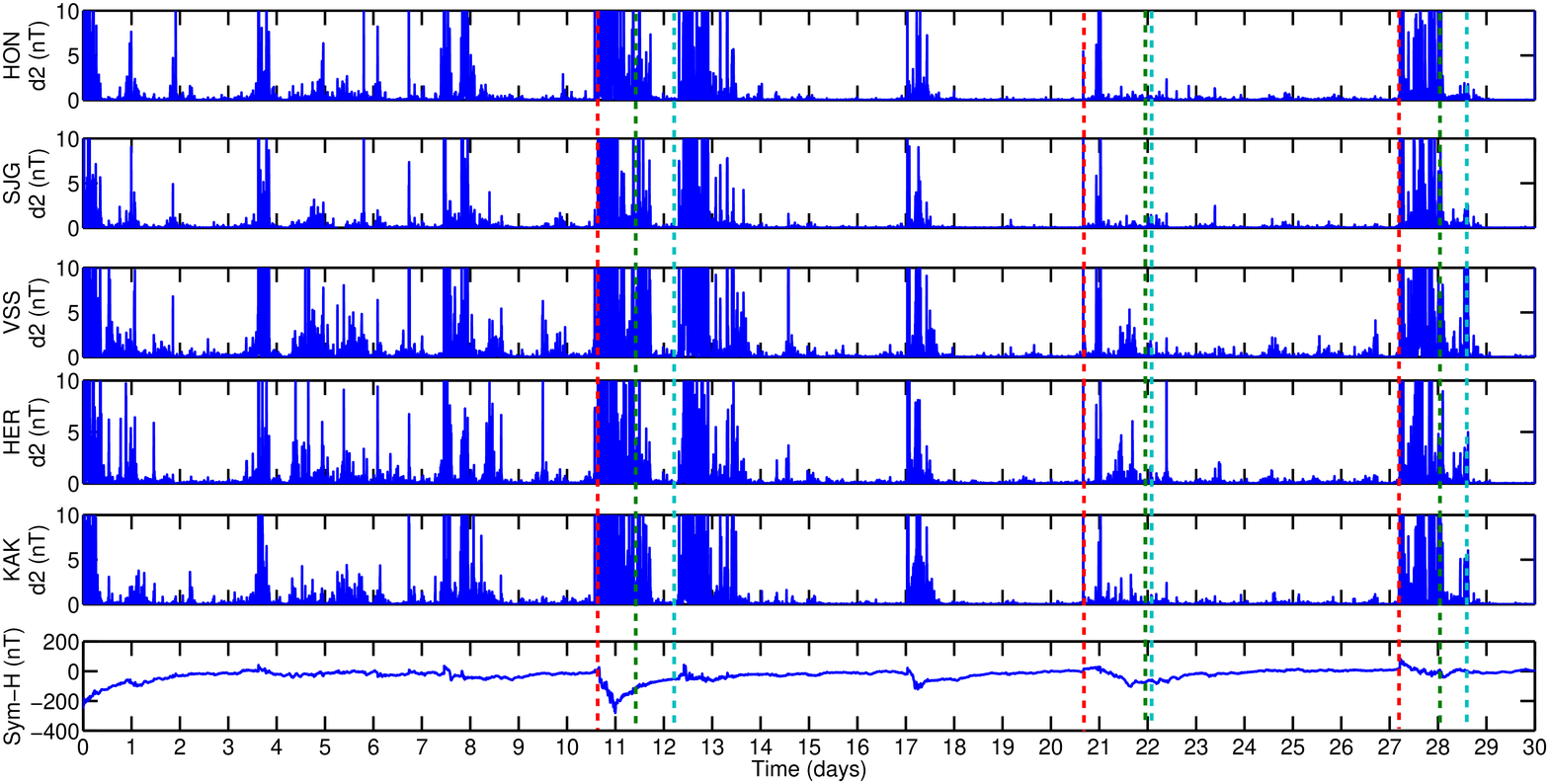} &
\includegraphics[width=10.5cm, height=6.5cm]{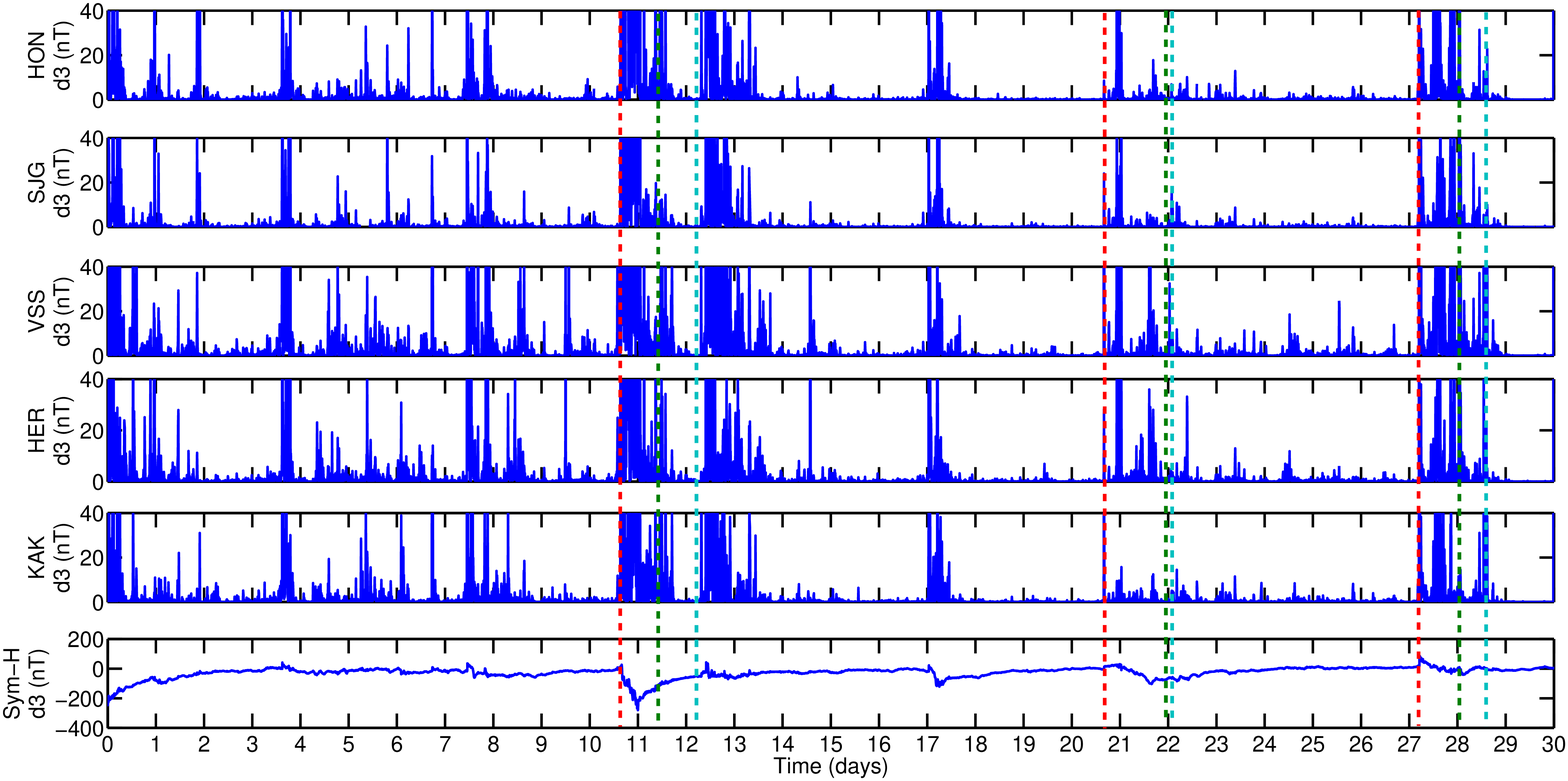}\\
\end{tabular}
}
\caption{Ground: (a) magnetograms and Sym-H index; and (b,c,d) the wavelet coefficients $(d^j)^2$ for $j=1,2,3$ and Sym-H index, respectively.}
\label{fig:Ground}
\end{figure}
\end{landscape}

Fig.~\ref{fig:Ground} shows four panels presenting (a) the ground magnetograms and (b--d) the first three decomposition levels, respectively.
Each panel displays, from top to bottom, the data or results for the magnetic stations of Honolulu (HON), San Juan (SJG), Vassouras (VSS), Hermanus (HER) and Kakioka (KAK), respectively, plus the Sym-H index.
We observe that the increase on the coefficient amplitudes during the sudden storm commencement (SSC), and also during the main phase of geomagnetic storm.
It is possible to notice that the wavelet coefficients presented an increase by the arrival time of the MCs when it reaches Earth's magnetosphere.
The MC candidates are displayed on Table~\ref{tabela:MCList} and \ref{tabela:MCListCane}.
All the magnetic stations showed similar pattern of magnetic behavior response to the storms with just few singularities in the three levels of wavelet decomposition.
As \cite{MendesMag2005} suggested, this behavior may be related to differences in the magnetic coordinates, type of magnetometer used, local time, ground conductivity and Sq currents effects.

\begin{landscape}
\begin{figure}[hbt]
\includegraphics[width=21cm]{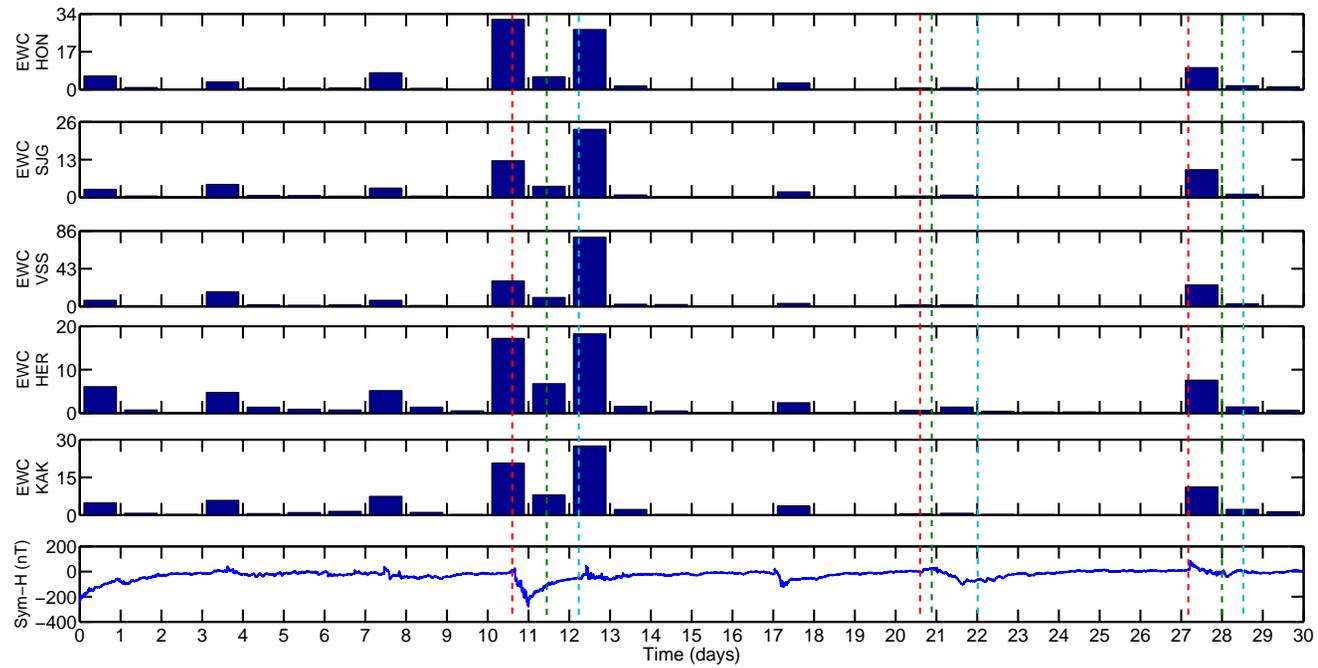}
\caption{The effectiveness wavelet coefficients on April, 2001 for the ground magnetograms.}
\label{fig:EWC_Ground}
\end{figure}
\end{landscape}

In Fig.~\ref{fig:EWC_Ground}, the first shock presented the higher EWCs, the same happened in Fig.~\ref{fig:EWC_ACE}.
Similar to Fig.~\ref{fig:EWC_ACE}, the second shock was not well detected in Fig.~\ref{fig:EWC_Ground}.
The third shock also presented the higher EWCs during the shock and its behavior was similar to Fig.~\ref{fig:EWC_ACE}.

Also, the EWCs shows an increase of values associated with the shocks listed on Table~\ref{tabela:MCListCane} and it was well-correlated to the arrival time of the ICMEs when it reaches Earth's magnetosphere.

\subsection{ACE and Ground data correlation analysis}

In the ACE parameter analysis, there was an increase of wavelet coefficient amplitudes and EWCs of the IMF components during shock and sheath region.
It can start a geomagnetic storm because during the passage of a sheath, the shock compression, turbulence, magnetic field draping or shock heliospheric current sheet may lead to southward magnetic fields \citep{Badruddin2002}.
As \Citet{Gonzalez1987} discussed, a geomagnetic storm occur when the interplanetary magnetic field turns southward and remains southward for an prolonged period of time.

\begin{figure}[hbt]
\centering
\includegraphics[width=11cm]{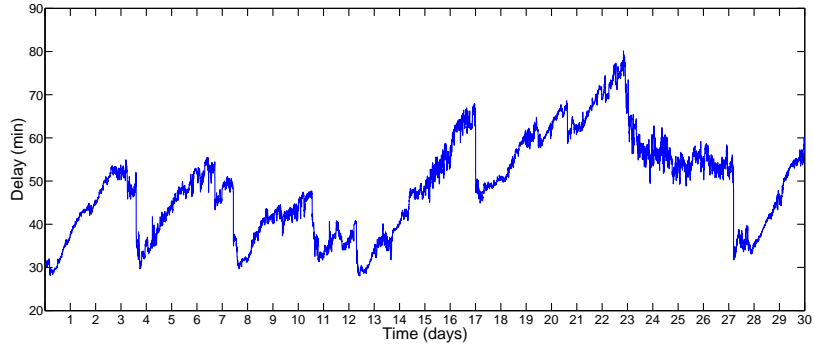}
\caption{ACE data: Minutely time delay of the solar events arrival on the Earth's magnetosphere for April 2001.}
\label{fig:ACEdelay}
\end{figure}

The ACE satellite is localized in the Lagrangian Point L1.
On April, 2001, it could detected solar events about 30 to 80 minutes before they arrived to the Earth's magnetosphere (see Fig.~\ref{fig:ACEdelay}).
If the DWT could be applied online then it could help to forecast a future geomagnetic disturbance.
Generally, models that predict geomagnetic activity work well at minimum solar activity.
Then, the capacity of the wavelet method to work correctly in high solar activity is an important advantage for its future use as a sophisticated space weather
tool.

In all the ground magnetograms analysis, the amplitude increase  of wavelet coefficients were more frequent during the sudden storm commencement (SSC), and also during the main phase of geomagnetic storm.
The SSC occurs due to sudden impulse due to the arrival of the interplanetary shock structure what generally coincides with the increase ram  pressure (initial phase) followed by the decrease of geomagnetic field (main phase) which indicates sustained southward interplanetary fields in the sheath region or/and during the MC propagation \citep{Gonzalez1994}.
It was possible to notice a simultaneous increase of EWC values on Fig.~\ref{fig:EWC_ACE} and \ref{fig:EWC_Ground}, respectively at, the shock and SSC regions; and the sheath and the main phase regions.
Unfortunately, the weighting the square wavelet coefficients means by hour (EWC hourly values) disguises the arrival delay of the shock and sheath region at the Earth's magnetosphere.

On the analyses of ground magnetograms, the highest coefficient amplitude were coincident in time, showing that the whole magnetosphere is globally affected, at least in the time resolution considered as discussed in \cite{MendesMag2005}.
The amplitude of the EWCs are related to geoeffectiveness of the solar events.
It is also worth to mention that the small amplitudes of the wavelet coefficients mean that the energy transfer process is smooth; while the large amplitudes indicated that were an impulsive energy injections superposed to the smooth background process.

Table~\ref{table:CorrelationApr2001} shows the determination coefficient matrix between the ten dataset parameters (Bx, By, Bz, density, Vx, HON, SJG, VSS, HER and KAK) used in our study.
The determination coefficient is defined as the square of the correlation coefficient.
One of the reasons to use the coefficient of determination instead of the correlation is to compute the statistics in order to determine the size or magnitude of the relation between two variables. 
It is interpreted as the percentage of variability of the response variable explained by the regression model. 
The correlation coefficient measures linear association.
Though in space geophysics both the determination and correlation coefficient are used \citep{Reiff1983}. 
In our case, we prefer to use the determination coefficient due to its interpretation of linear regression.

\begin{table}[htp]
\scriptsize{
       \caption{Determination coefficient matrix between the ten parameters used in this study of solar wind and magnetogram datasets.}
       \centering
         \begin{tabular}{c| c c c c c c c c c c }
         \hline
         & $B_x$ & $B_y$ & $B_z$ & $n_p$ & $V_x$  & $HON$& $SJG$&
$VSS$& $HER$& $KAK$\\
\hline
 $B_x$ & 100.   &       &       &       &       &       &       &       &       &   \\
 $B_y$ & 90.82   &100.   &  &  &   &   &   &    &    & \\
 $B_z$ & 92.54   &90.82   &100.   &   &   &   &  &    &     & \\
$n_p$ &  18.66  &21.34   &25.40   &100.   &  &   &    &     &    & \\
$V_x$ & 53.14    &33.29   &42.51   &0.57   &100.   &   &  &    &    & \\
$HON$ & 52.71    &58.37   &46.65   &10.50  &22.56   &100.   &   &    &    & \\
$SJG$ & 29.92    &29.37   &26.21   &4.84    &27.14   &83.91   &100.   &    &    & \\
$VSS$ & 17.81    &17.56   &15.29   &2.25    &21.34   &72.08   &97.02   &100.   &   & \\
$HER$ & 45.97   &48.02   &40.96    &8.41    &27.56   &95.45   &89.49   &79.92   &100.   & \\
$KAK$ & 40.45   &41.34   &35.64   &7.56   &28.94   &92.92   &95.84   &88.55   &97.22   &100.\\
         \hline
        \end{tabular}
}
      \label{table:CorrelationApr2001}
    \end{table}

As we mentioned above, the larger values of determination coefficient between the five magnetic stations reinforces the statement that whole magnetosphere is globally affected by the solar disturbances, see Table~\ref{table:CorrelationApr2001}.
At the same time, VSS station presented the lower values of determination coefficient between the five magnetic stations.
This behavior of VSS can be explained by its localization which is in a region very peculiar, under the influence of the SAMA that is characterized by a global minimum in the Earth's total magnetic field intensity.
As suggested by \citet{PintoGonzalez1989}, the SAMA region can be compared to the auroral region mainly during geomagnetic storms due to the development of some current-driven plasma instabilities.
This peculiar behavior of VSS was also observed by \cite{MendesMag2005} in their study using several magnetic stations and geomagnetic storms.

Also, it is possible to notice larger values of determination coefficient between the Bx, By and Bz solar wind parameter is expected because their are the magnetic component of the same magnetic vector.
However, between the solar wind parameters and the five magnetic stations the determination coefficients present lower values due to the amplitude of the wavelet coefficients are related to the propriety of the db2 wavelet to reproduce linear function locally, this means that it is sensible to local variations on the solar parameters.
In other words, not every fluctuation of the solar parameters has enough energy to disturbed the Earth's magnetosphere but it is detected by DWT analysis.

\section{Conclusions}
\label{Final Remarks}

In this work, we investigate the magnetic disturbances that occurred on April, 2001 sing a new methodology based in the amplitude of the discrete wavelet coefficients applied to solar wind and magnetogram datasets.
We select the solar parameters detected by the ACE satellite and the H-component of magnetic field measured by ground magnetometers at HON, SJG, VSS, HER and KAK.
The magnetic station of VSS were selected as a case study to verify the geomagnetic responses to this event under the region of the SAMA influence.

In response to the disturbances on April, 2001, the results show that the increase of wavelet coefficient amplitudes associated with the solar wind parameters and H-component were well correlated with the shock and sheath regions.
In the ground magnetograms analysis, the amplitude increase of wavelet coefficients were well correlated with the sudden storm commencement (SSC) and the main phase of geomagnetic storm.
These phase of the geomagnetic are associated with the arrival of the shock and sheath regions on the Earth's magnetosphere, respectively.
Our results also show that the VSS station presented a peculiar behavior that could be explained due to it is located under the influence of the SAMA.

The effectiveness wavelet coefficient (EWC) methodology facilitates the visualization and the analysis of the wavelet coefficients because it only have one decomposition level.
This new approach suggests a new representation of the results.
The previous studies of \cite{MendesMag2005} and \cite{MendesdaCostaetal:2011}, employed three decomposition levels.
Our methodology could be used in a semi-automatic way to characterize the solar wind and ground magnetic disturbances.
It could be important to improve the knowledge of the peculiarities and the effects produced by solar events on the Earth's magnetosphere and further it could be implemented to real-time analysis for forecast space weather scenarios.
And also, it could enable us to study certain aspects of processes involved in the solar wind-magnetosphere interaction. 

\section{Acknowledgments}
V. Klausner wishes to thanks CAPES for the financial support of her PhD (CAPES -- grants 465/2008) and her Postdoctoral research (FAPESP -- 2011/20588-7).
A. O. González thanks CAPES and CNPq (141549/2010-6) for his PhD scholarship.
This work was supported by CNPq (grants 309017/2007-6, 486165/2006-0, 308680/2007-3, 478707/2003, 477819/2003-6, 382465/01-6), FAPESP (grants
2007/07723-7) and CAPES (grants 86/2010-29, 0880/08-6, 86/2010-29, 551006/2011-0, 17002/2012-8). 
Also, the authors would like to thank the NOAA and the INTERMAGNET programme for the datasets used in this work.

\end{document}